\documentclass[5p,twocolumn,times,number]{elsarticle}

\usepackage{graphicx}
\usepackage{amsmath}   
\usepackage{lipsum}
\usepackage{mathtools}
\usepackage{lineno,hyperref}
\usepackage{graphicx}
\usepackage{subfig}
\usepackage{multirow} 
\usepackage{rotating} 
\usepackage{adjustbox}
\usepackage{pbox} 
\usepackage{array}
\usepackage{float}
\usepackage{makecell}
\usepackage[utf8]{inputenc}
\usepackage{longtable}
\usepackage{hyperref}

\begin{document}

\begin{frontmatter}

\title{Study of radon reduction in gases for rare event search experiments}

\author[affliation1]{K. Pushkin\corref{cor1}}
\ead{kpushkin@umich.edu}
\cortext[cor1]{Corresponding author}
\author[affliation1]{C. Akerlof\corref{cor2}}
\author[affliation1]{D. Anbajagane\corref{cor2}}
\author[affliation2]{J. Armstrong\corref{cor2}}
\author[affliation1]{M. Arthurs\corref{cor2}}
\author[affliation2]{J. Bringewatt\corref{cor2}}
\author[affliation2]{T. Edberg\corref{cor2}}
\author[affliation2]{C. Hall\corref{cor2}}
\author[affliation1]{M. Lei\corref{cor2}}
\author[affliation1]{R. Raymond\corref{cor2}}
\author[affliation1]{M. Reh\corref{cor2}}
\author[affliation1]{D. Saini\corref{cor2}}
\author[affliation1]{A. Sander\corref{cor2}}
\author[affliation1]{J. Schaefer\corref{cor2}}
\author[affliation2]{D. Seymour\corref{cor2}}
\author[affliation2]{N. Swanson\corref{cor2}}
\author[affliation1]{Y. Wang\corref{cor2}}
\author[affliation1]{and W. Lorenzon\corref{cor2}}

\address[affliation1]{Randall Laboratory of Physics, University of Michigan, Ann Arbor, MI 48109-1040, USA}
\address[affliation2]{Department of Physics, University of Maryland, College Park, MD 20742-4111, USA}

\begin{abstract}
The noble elements, argon and xenon, are frequently employed as the target and event detector for weakly interacting particles such as neutrinos and Dark Matter. For such rare processes, background radiation must be carefully minimized. Radon provides one of the most significant contaminants since it is an inevitable product of trace amounts of natural uranium. To design a purification system for reducing such contamination, the adsorption characteristics of radon in nitrogen, argon, and xenon carrier gases on various types of charcoals with different adsorbing properties and intrinsic radioactive purities have been studied in the temperature range of 190-295 K at flow rates of 0.5 and 2 standard liters per minute. Essential performance parameters for the various charcoals include the average breakthrough times ($\tau$), dynamic adsorption coefficients ($\textit{$k_a$}$) and the number of theoretical stages ($\textit{$n$}$). It is shown that the $\textit{$k_a$}$-values for radon in nitrogen, argon, and xenon increase as the temperature of the charcoal traps decreases, and that they are significantly larger in nitrogen and argon than in xenon gas due to adsorption saturation effects. It is found that, unlike in xenon, the dynamic adsorption coefficients for radon in nitrogen and argon strictly obey the Arrhenius law. The experimental results strongly indicate that nitric acid etched Saratech is the best candidate among all used charcoal brands. It allows reducing total radon concentration in the LZ liquid Xe detector to meet the ultimate goal in the search for Dark Matter. 

\end{abstract}

\begin{keyword}
Radon \sep Noble Gases \sep Dual Phase Detector \sep Time-Projection Chamber \sep Dark Matter \sep Neutrinoless Double Beta Decay

\end{keyword}

\end{frontmatter}

\section{Introduction}

Modern rare-event search experiments require low-radioactivity Time Projection Chambers (TPCs) to achieve high detection sensitivities.  Noble gases such as argon (Ar) and xenon (Xe) are well-suited as target media for Dark Matter (DM) and Neutrinoless Double Beta Decay (NDBD)\cite{Akerib2017, EXOcollab, Guiseppe2017}.  They are both excellent scintillators.  Both are easily ionized by particles and can be easily purified of electronegative impurities \cite{Aprile_book}  to achieve efficient charge transport.  The challenge of every DM and NDBD experiment is to suppress radioactive backgrounds.  Natural Ar and Xe do not have intrinsic long-lived isotopes but, during the production cycle, Xe can be contaminated with $^{85}$Kr from the atmosphere. $^{39}$Ar is produced in the atmosphere by cosmic rays scattering from $^{40}$Ar. Both radioactive isotopes decay primarily by $\beta$-emission and their presence in detectors may limit ultimate sensitivities.  Fortunately online distillation systems can be employed to remove these radioactive isotopes \cite{Wang2014}. Radon ($^{222}$Rn), with a half life of 3.8 days, is another isotope that must be eliminated from TPC detectors. $^{222}$Rn is a daughter of $^{238}$U and is continuously supplied from detector components (e.g., cables, feedthroughs).
\par
In the past charcoals have been used in low background experiments (Borexino \cite{Pocar2003}, XMASS \cite{Abe2012}, SNO+  \cite{Golightly2008} and CUORE \cite{Benato2018}) to reduce  $^{222}$Rn in the detectors by trapping the $^{222}$Rn in charcoal  long enough for it to decay but, to the best of our knowledge, available charcoal adsorbents have not been systematically studied.  A gas system was fabricated to study $^{222}$Rn reduction methods using commercial charcoal brands: Calgon Carbon (OVC 4x8) \cite{Calgon}, Shirasagi (G2x4/6-1) \cite{Shirasagi}, Saratech (Bl{\"u}cher GmbH.), nitric acid (HNO$_3$) etched Saratech (Bl{\"u}cher GmbH.) \cite{Saratech} and Carboact (Carboact International) \cite{Carboact}. The studies were performed within the scope of the LUX-ZEPLIN (LZ)  DM experiment \cite{Akerib2018}. The physical conditions of charcoals ranged between 190 K and 295 K for $^{222}$Rn in nitrogen (N$_{2}$), Ar and Xe as carrier gases with flow rates of 0.5 and 2 standard liters per minute (slpm).  

\section{Apparatus}

\subsection{Gas system}

$^{222}$Rn adsorption on different types of charcoals was measured while entrained in N$_{2}$, Ar, and Xe carrier gases. A schematic view of the $^{222}$Rn reduction system is shown in Figure 1. The radon reduction system was designed and constructed at the University of Michigan. 

\begin{figure*}[h!]%
\centering
\includegraphics[height=2.5in]{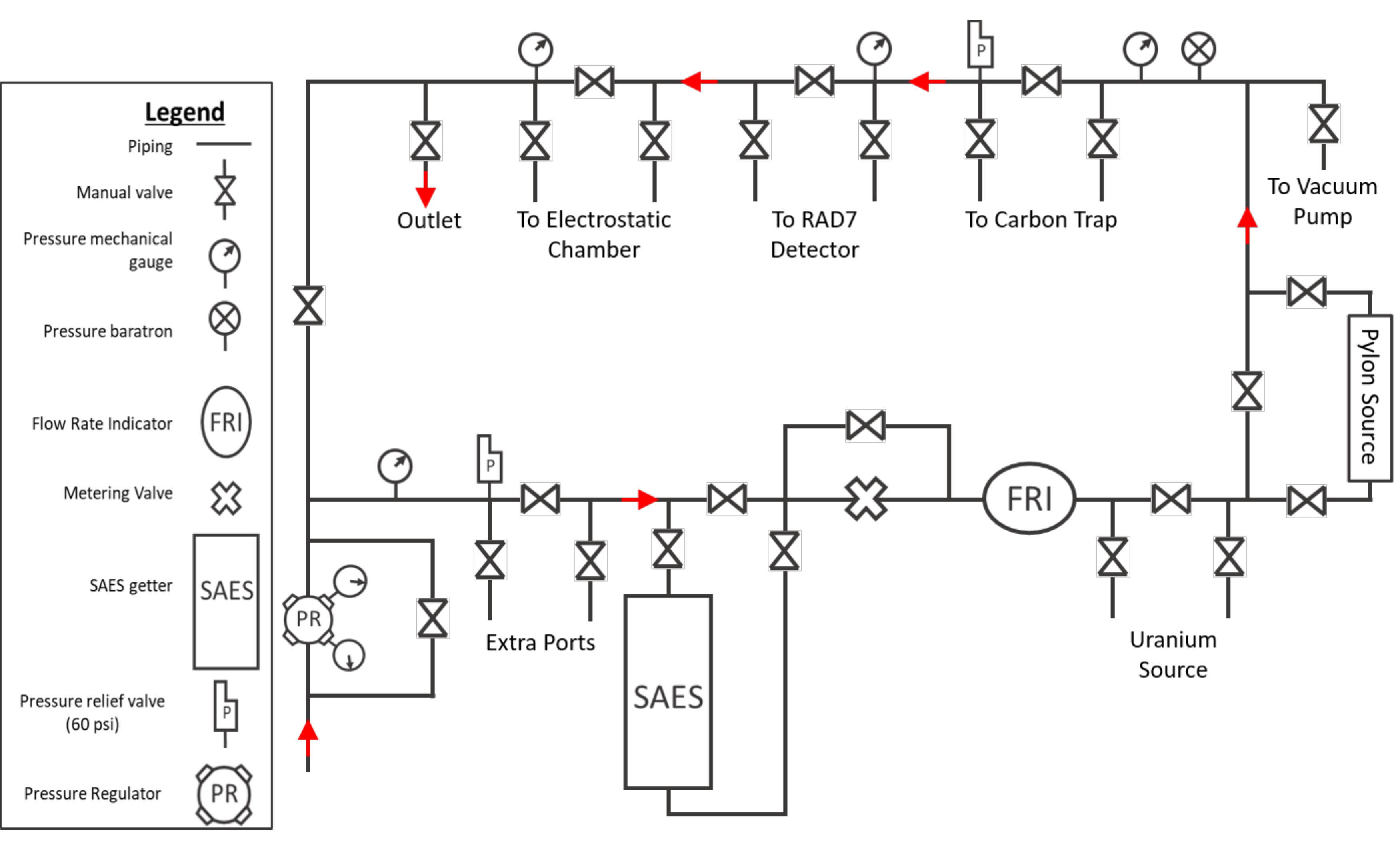}
\caption{Schematic view of the $^{222}$Rn reduction and evaluation gas system.} 
\end{figure*}%

The plumbing assemblies were welded (HE Lennon Inc., Swagelok, Michigan) \cite{HELennon} from 1/4" ultra-high vacuum (UHV) pipes and valves. After completion, the system was helium leak checked using a residual gas analyzer (RGA100, SRS), and no leaks were detected within the detection limits (5$\times$10$^{-14}$ Torr). Water vapor and other gas impurities may significantly affect the performance of a charcoal trap. Therefore, only boiled-off N$_{2}$ and Ar gases as well as research purity grade Xe gas were used for the measurements. Moreover, Xe and Ar gases were continuously purified during the measurements with a SAES high temperature getter (Model: PS3-MT3-R-1) composed of zirconium to remove oxygen, water vapor, nitrogen and other impurities \cite{SAES}. The getter bed operating temperature was in the range of 350-400 $^{0}$C. When pure N$_{2}$ was used, the getter was turned off and bypassed to prevent exothermic reaction leading to irreversible damage, and possibly explosion, to the zirconium cartridge. Before commencement of each set of measurements with carrier gases and charcoals, the moisture content of the trap was purged with boiled-off N$_{2}$ gas for 12 hours. The charcoal trap was baked at 100 $^{0}$C and evacuated with scroll and turbo-molecular pumps (Agilent Technologies). The charcoal trap and the UHV gas system were evacuated to at least 10$^{-5}$ Torr for 12 hours before the measurements began.
\par
The experimental procedure was as follows. At the beginning of each measurement the gas was flowed through the charcoal trap and two independent radon detectors connected in series to evaluate radon background levels in the gas system and in the investigated charcoal trap. Uranium ores were initially used as $^{222}$Rn source, but their activity was very weak. Therefore, they were replaced by a Pylon-1025 source (Electronics Development Company, Ltd.) containing dry radium ($^{226}$Ra) with an activity of 103 kBq and encapsulated in an aluminum cylinder to prevent its leakage \cite{Pylon}. The carrier gas was diverted through the source for up to 3 minutes resulting in an injection of a sharp, short pulse of $^{222}$Rn into the charcoal trap. The carrier gas continued flowing through the trap for the duration of each individual measurement. The gas flow rate was controlled with a UHV gas regulator and a metering valve. It was measured with a UHV mass flow meter (Model:179A01314CR3AM, MKS) \cite{MKS}. The precision and accuracy of the flow meter were 0.2\% and 1\% of the full scale, respectively. The pressure in the gas system was regulated using a check-valve installed at the outlet port. The mass flow meter was operated with a single channel power supply readout (Model: 246C, MKS) \cite{MKS}. The gas system included mechanical gas pressure gauges and a baratron absolute pressure transducer with an analog read-out before and after the trap to measure the gas pressure. The measurements were made at different gas pressures to study their impact on $^{222}$Rn adsorption on charcoals. The gas system was equipped with pressure relief valves (PRV) rated to 60 pounds per square inch (psi). When average $^{222}$Rn breakthrough times were measured in Ar and N$_{2}$ carrier gases, the gases were not recovered in view of their low cost. The system handled 16 kg of Xe gas contained in two aluminum cylinders with volumes of 30 l each. The xenon gas was cryogenically transferred between the two cylinders using liquid N$_2$. The mass of the Xe gas was measured with tension load cells (FL25-50 kg, Forsentek, China) \cite{Forsentek}.

\subsection{Vacuum-jacketed cryostat and charcoal traps}

In order to select the optimal radon adsorbent, different types of charcoals were investigated. Charcoals were contained in modified conflat UHV vessels (Kurt J. Lesker, USA) \cite{Lesker}. The dimensions of the vessels were 3.4 cm and 6.4 cm in diameter and 12.6 cm and 35 cm in length with corresponding volumes of 0.1 l and 1.1 l, respectively. Charcoals Calgon OVC 4x8 (Calgon, 50 g, 0.1 l), Shirasagi (G2x4/6-1, 45 g, 0.1 l), Saratech (70 g, 0.1 l; 650 g, 1.1 l), HNO$_3$ etched Saratech (650 g, 1.1 l) and Carboact (241 g, 1.1 l) were selected for their different properties such as porosity, density, surface area, radioactive background as well as relative cost. Before assembly, the traps and their UHV components were thoroughly cleaned in methanol. All charcoals, except specially treated etched Saratech, were then rinsed with deionized (DI) water. The traps were equipped with fine stainless steel (SS) meshes as well as  60 $\mu$m UHV filter gaskets on the inlet and outlet to prevent escape of fine charcoal grains and dust into the main volume of the gas system. The pressure difference between the inlet and outlet of the charcoal trap was measured to be less than 1 psi. The charcoal bed layers were compressed with a SS spring to maintain stable and uniform packing. The meshes and springs were cleaned in an ultrasonic bath with methanol prior to assembly. The charcoal traps were equipped with calibrated platinum RTD PT100 (Omega) temperature sensors \cite{Omega} embedded in the charcoal layers. The temperatures of the inner volume in the vacuum-jacketed cryostat and the inner volume of the charcoal trap agreed within 0.5\%. The vacuum-jacketed cryostat was well suited (Cryofab, Cryogenic equipment, USA) \cite{Cryofab} to store charcoal traps and perform studies at various temperatures. The inner height and diameter of the cryostat were 132 cm and 30 cm, respectively. The cryostat was filled with 57 l (87 kg) of Novec-7100 (C$_4$F$_9$OCH$_3$, 3M, Engineered fluid)  \cite{Novec} to provide uniform and efficient cooling of the charcoal traps. The freezing and boiling temperatures of Novec-7100 fluid are 135 K and 334 K, respectively. The top lid of the cryostat was insulated from the trap and cryofluid by a 30 cm thick polystyrene layer to reduce thermal conduction. The cryostat was equipped with a baratron pressure transducer, a PRV, a manual gas valve, a solenoid valve for the engineered fluid and an outlet port for the refrigerator cooling probe. A photograph of the vacuum-jacketed cryostat and its cutaway view with a small charcoal trap (0.1 l) inside are shown in Figure 2 (a, b).  

\begin{figure}[!tbp]
 \centering
 \subfloat[]{\includegraphics[width=0.35\textwidth]{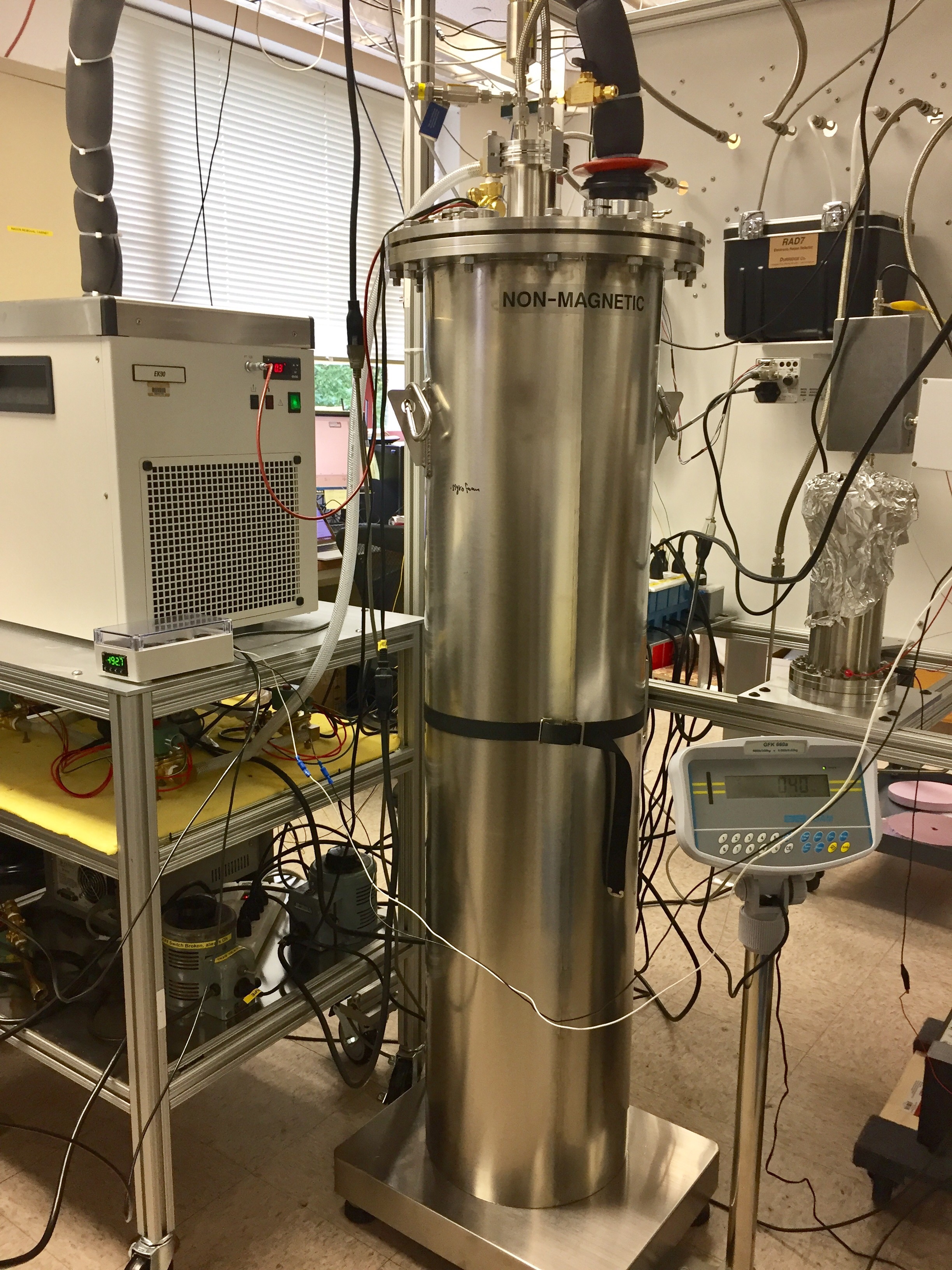}\label{fig:f1}}
 \vspace{0.00001cm}
 \hfill
 \subfloat[]{\includegraphics[width=0.35\textwidth]{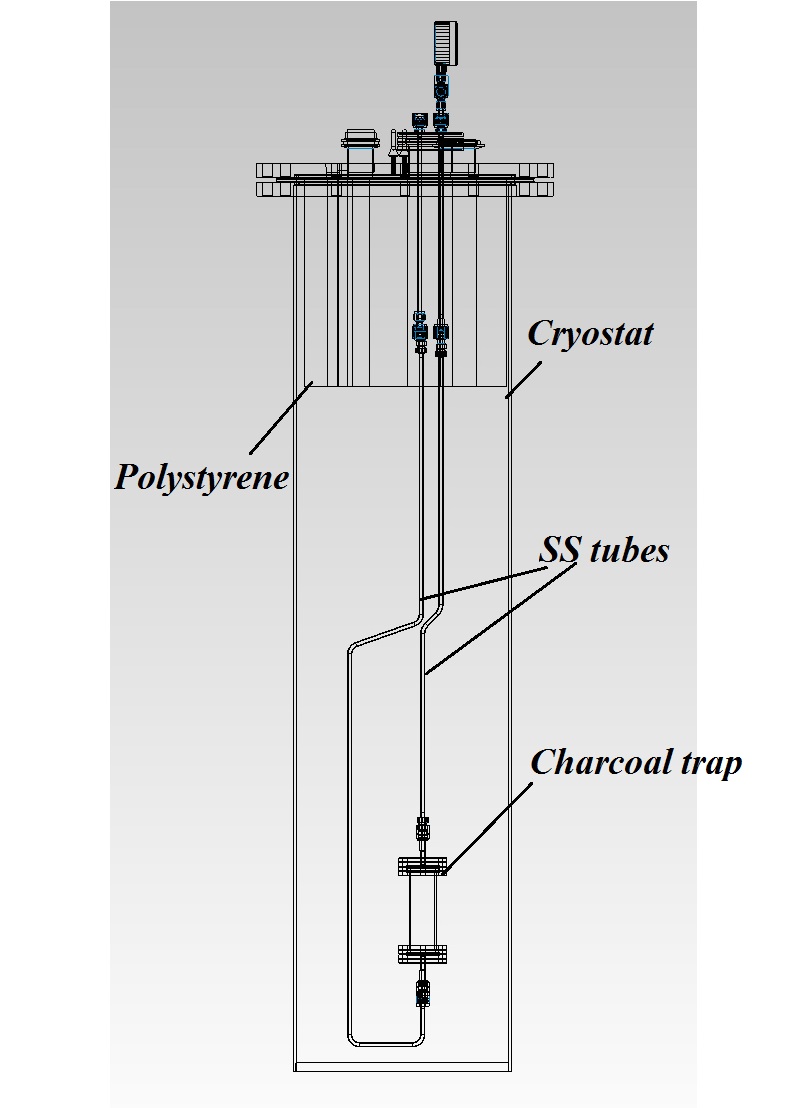}\label{fig:f2}}
 \caption{Photograph of the vacuum-jacketed cryostat (a) and a cutaway view of the cryostat with a 0.1 l trap (b).}
\end{figure}

An EK-90 immersion cooler (Thermo Fisher Scientific, USA) \cite{EK90} was used to cool the charcoal traps to 190-273 K. The immersion cooler was equipped with a digital temperature LED display and a calibrated platinum RDT PT100 sensor to monitor the temperature of the cooled fluid. Moreover, the cryostat was equipped with a 500 W heater to warm the Novec-7100 fluid and  heating tape to bake the charcoal traps. A quad chamber diaphragm pump (Grainger, USA) with a manifold consisting of electric solenoid valves and Tygon PVC pipes was used to siphon Novec-7100 fluid between the cryostat and the sealed storage drum. To monitor adsorption of the carrier gases on the charcoal during measurements, the cryostat was placed on an electronic scale (GFK 660a, Adam) with a maximum capacity of 300 kg and an accuracy of 0.02 kg.

\subsection{General characteristics of the investigated charcoals}

The description of the charcoals used in this work is shown in Table 1. Both the radiation purity of the charcoals and their adsorbing characteristics are crucial for their applications in low background experiments. $^{222}$Rn emanation from the charcoals may significantly contribute to radioactive background in TPC detectors. Hence, for the development of a charcoal trap, both the $^{222}$Rn emanation from within the detector and from the charcoal itself must be taken into account. 


\begin{table*}[h!]
\centering
\begin{tabular}{c|c|c|c|c}
\hline
\text{Charcoal} & \text{Shape} & \thead{\text{Bulk density} \\ $(g/cm^3)$} & \thead{\text{Surface area} \\ $(cm^2/g)$} & \thead{\text{Size} \\ $(cm)$} \\
\hline
Calgon OVC 4x8 & Flake & 0.45 & 1.1$\times$10$^7$ & 0.3 - 0.8 \\
Shirasagi G2x4/6-1 & Cylinder & 0.40 - 0.47 & 1.24$\times$10$^7$ & 0.3 - 0.5 \\
Saratech & Sphere & 0.60 & 1.34$\times$10$^7$ & 0.05 \\
Carboact & Fragmented & 0.28 & 8$\times$10$^6$ - 1.2$\times$10$^7$ & 0.01 - 0.4
\end{tabular}
\caption{List of charcoals and their properties (the physical properties of HNO$_3$ etched Saratech remained unchanged relative to those ones of regular Saratech).}
\end{table*}

The radon emanation rates of the charcoal samples were measured with a counting facility developed for the LZ experiment and were carried out at the University of Maryland. The facility features conflat (CF) vessels for hosting the charcoal samples during radon emanation, a radon trapping gas panel, and a low-background electrostatic radon counter (ESC) similar to that shown in Figure 3. 
Each charcoal sample was weighed, rinsed with DI water five to ten times, and loaded into a CF vessel. The carbon was compressed slightly on both ends with springs and secured with stainless steel perforated meshes and polyester felt. The carbon was dried for no less than 12 hours with a N$_2$ gas purge while baking at a temperature between 100 $^{0}$C and 140 $^{0}$C. After connecting the vessel to the $^{222}$Rn trapping panel, the charcoal was prepared for emanation by purging with radon-free helium (He) gas at a temperature of 140 $^{0}$C. This removed all relic $^{222}$Rn from the charcoal. To determine the amount of He carrier gas required for this purge, a preparatory measurement was performed on each sample using a $^{222}$Rn source and a specified He flow rate calibrated with a bubble flow meter. 

After a given emanation period, typically one week, the charcoal was heated to 140 $^{0}$C and the emanated $^{222}$Rn was recovered by purging the charcoal again with radon-free He. To collect the recovered $^{222}$Rn, the He purge was directed through a liquid N$_2$ cold trap filled with copper beads and then pumped out of the system. After warming the copper trap to room temperature, the trapped $^{222}$Rn was transferred into the electrostatic counter for measurements of He at atmospheric pressure. The trapping efficiency was shown to be near 100\% by calibration measurements. The absolute efficiency of the counter was determined to be (24$\pm$4)\% by measuring a calibrated radon source purchased from Durridge \cite{Durridge}. A second calibrated source provided by researchers at Laurentian University provided an independent cross check that agreed within the measurement uncertainty. In addition, ion drift simulations of the counter were performed. These confirmed that the measured counter efficiency is reasonable. The simulations rely upon previous measurements of the ion charge fraction in gases such as N$_2$ and He \cite{Andersen1997, Howard1991}.

To search for a temperature dependence of the emanation rate, we allowed each charcoal sample to emanate at temperatures of 20 $^{0}$C, 80 $^{0}$C, and 140 $^{0}$C. No temperature dependence was observed over this range for all of the carbon samples. This suggests that either the $^{222}$Rn is produced at the surface of the carbon grains, or else the characteristic time for a $^{222}$Rn atom to diffuse out of the bulk at these temperatures is short compared to the radon half-life. A solution of the diffusion equation in a homogenous spherical geometry indicates that the egress fraction due to diffusion should remain near unity at these temperatures. An activation energy of 29,000 J/mol$\cdot$K, and a diffusion constant below $10^{-3}$ cm$^2$/s was assumed.

The mass measurement of the charcoal was subject to systematic uncertainty. Each charcoal sample was weighed upon receipt from the vendor and prior to washing in DI water. After washing, the CF vessel was filled with the wet charcoal, and the vessel was baked and purged overnight with N$_2$ gas to dry as explained above. The portion of wet charcoal which did not fit in the vessel was set aside and allowed to air dry for one week. Its weight was then measured and subtracted from the total mass to determine the mass of charcoal loaded into the vessel. The amount subtracted was less than 2\% of the total except for Calgon OVC (8\%). The results of the specific activity for natural and synthetic charcoals reported in Table 2 were obtained by dividing the observed emanation rate by the mass according to the procedure. A blank measurement of the empty CF vessel was subtracted from each result. In Table 2 we combine all measurements for a given sample, regardless of emanation temperature, using either a weighted average or a maximum likelihood method. The uncertainties reported in Table 2 are statistical only. A 16\% systematic uncertainty due to the ESC efficiency applies to all results.

Among the samples tested, the synthetic carbons had the lowest specific activity. Carboact was found to have the very lowest activity, followed by a sample of Saratech adsorbent that was soaked in a 4 M solution of ultra-pure HNO$_3$ acid and rinsed in DI water. We refer to this material as HNO$_3$ etched Saratech. 363 grams of Carboact and 796 grams of etched Saratech were employed for these measurements. To increase the statistical power, three Carboact and three etched Saratech emanation runs were performed. Regular (HNO$_3$ unetched) Saratech was also measured and was found to have a specific activity several times higher than the etched material, indicating the efficacy of the etching process for removing the traces of $^{238}$U. The bulk density of etched Saratech was measured after the etching procedure and remained unchanged.

\begin{table*}[h!]
\centering
\begin{tabular}{c|c|c|c}
\text{Charcoal} & \thead{\text{Specific activity} \\ $(mBq/kg)$} & \thead{\text{Price} \\ $(USD/kg)$} & \text{References}\\
\hline
Calgon OVC 4x8 & 53.6 $\pm$ 1.3 & 6 & This work\\
Shirasagi G2x4/6-1 & 101.0 $\pm$ 8.0 & 27 & This work\\
Saratech & 1.71 $\pm$ 0.20 & 35 & This work\\
HNO$_3$ etched Saratech & 0.51 $\pm$ 0.09 & 135 & This work\\
Carboact & 0.23 $\pm$ 0.19 & 15,000 & This work\\
Carboact & 0.33 $\pm$ 0.05 & 15,000 & \cite{Rau2000}
\end{tabular}
\caption{Radon emanation screening from charcoals and their approximate prices in 2017. Uncertainties are statistical only and are reported at 68\% C.L. The systematic uncertainty on the specific activity for the Calgon OVC is 8\% due to the mass measurement.}
\end{table*}

\subsection{$^{222}$Rn electrostatic detectors}

A commercial radon detector (RAD7, Durridge, Radon Instrumentation) with real time monitoring and spectrum analysis was used to measure $^{222}$Rn average breakthrough times in N$_{2}$ and Ar carrier gases. Since the RAD7 detector is not UHV rated, a separate in-house ESC was constructed  to measure radon breakthrough times in Xe gas. The detector consisted of an UHV conflat vessel with a volume of about 2 l and a 18x18 mm$^2$ Si-PIN photodiode (S3204-09, Hamamatsu, 0.3 mm in depletion thickness, unsealed) \cite{Hamamatsu}. The schematic view of the ESC with electronics is shown in Figure 3.

\begin{figure}[h!]%
        \centering
        \includegraphics[height=2.5in]{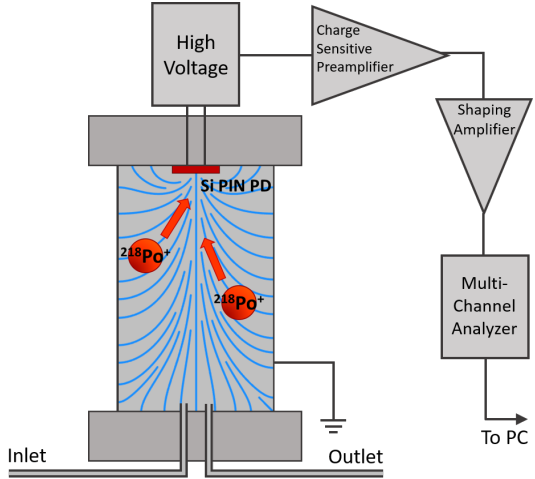}
        \caption{Schematic view of the ESC. The ancillary electronics is described in the text.} 
\end{figure}%

The detector performance is based on the electrostatic collection of $^{222}$Rn daughter nuclei \cite{Mitsuda2003, Mamedov2011}. As $^{222}$Rn atoms in the carrier gas enter the detector, they may decay and produce $^{218}$Po atoms. 88\% of the time, these $^{218}$Po atoms form positive ions which, in turn, are attracted towards the surface of the Si-PIN photodiode under the strong electric fields created within the vessel. At this point, subsequent alpha emission generates particles with kinetic energies of 6.00 MeV \cite{Hopke1996}. The Si-PIN photodiode was mounted on a high voltage feedthrough which was electrically isolated from the grounded stainless steel vessel. A high voltage (HV) divider was built to supply up to -6 kV \cite{Mitsuda2003}. The HV divider was physically separated from a charge sensitive preamplifier (CSP, Cremat, CR110) to prevent current leakage. The reverse voltage (-800 V) was applied through the HV divider to  the anode of the Si-PIN photodiode resulting in maximum collection efficiency of  $^{218}$Po ions. The reverse voltage difference between the cathode and the anode of the Si-PIN photodiode was below the maximum allowed reverse voltage ($\leq$100 V) for this particular Si-PIN photodiode. The charge signals were sent from the cathode to the CSP followed by a shaping amplifier (SA, Tennelec, TC243) with a shaping time of 1 $\mu$s. These amplified pulses were subsequently digitized by a multi-channel analyzer (MCA8000D, Amptek). The detection sensitivity was measured for both the RAD7 and in-house detectors using the Pylon source. The detection sensitivity for RAD7 was found to agree with its specification within statistical uncertainties for $^{218}$Po alpha-peaks \cite{RAD7} and was (6.34$\pm$0.02)$\times$10$^{-3}$ cpm/Bq/m$^3$. The detection sensitivity for our in-house ESC detector was found to be greater than for RAD7. It was measured to be (21.80$\pm$0.06)$\times$10$^{-3}$ cpm/Bq/m$^3$ for $^{218}$Po alpha-peaks. Furthermore, the in-house ESC detector can be operated at high gas pressures ($\approx$ 2 atm) while the operating gas pressure for RAD7 is not recommended to exceed 1 atm.

\subsection{Slow control system}

A slow control system was developed to monitor and control the important functions of the system. Parameters such as gas pressure, Novec-7100 fluid temperature, gas flow rates, charcoal temperature, Xe gas mass, room temperature, and room humidity were continuously monitored and recorded during operations. The values were archived in a SQL database and displayed on a  remotely monitored webpage. Additional automatic control features were introduced to control the gas system remotely. A liquid cryogenic control system was constructed for the $^{222}$Rn breakthrough time measurements in Xe gas to facilitate long term and safe measurements.

\section{Experimental results and discussion}

\subsection{Measurements of $^{222}$Rn adsorption characteristics on various charcoals in N$_2$, Ar, and Xe carrier gases}

$^{218}$Po and $^{214}$Po spectra were measured with time intervals from 3 to 15 minutes after a $^{222}$Rn spike was injected into the charcoal trap. Only $^{218}$Po was of interest since it is the first progeny decay product ("new radon") of the $^{222}$Rn decay chain. The $^{218}$Po peaks were fitted with an analytical function for alpha particle spectra where the integral-area of the peak was determined according to  \cite{Bortels1987} 

\begin{equation}
f({x,\mu,\sigma,\nu})=\frac{A}{2\nu}e^{{(\frac{x-\mu}{\nu}} + {\frac{\sigma}{2\nu^2})}}erfc\bigg(\frac{1}{\sqrt2}\bigg[\frac{x-\mu}{\sigma}+\frac{\sigma}{\nu}\bigg]\bigg),
\end{equation}

where $x$ is the channel number, $\textit{A}$ is the peak area, $\mu$ is the mean of the Gaussian probability-density function, $\sigma$ is the standard deviation and $\nu$ is the parameter of the normalized left-sided exponential function. A typical radon daughter pulse height spectrum is shown in Figure 4. Figure 5 displays examples of elution curves measured in Ar carrier gas through three types of charcoals Calgon OVC 4x8, Shirasagi, and Saratech) contained in a 0.1 l trap at a temperature of 295 K and at atmospheric pressure. 

\begin{figure}[h!]%
\centering
\includegraphics[height=3in]{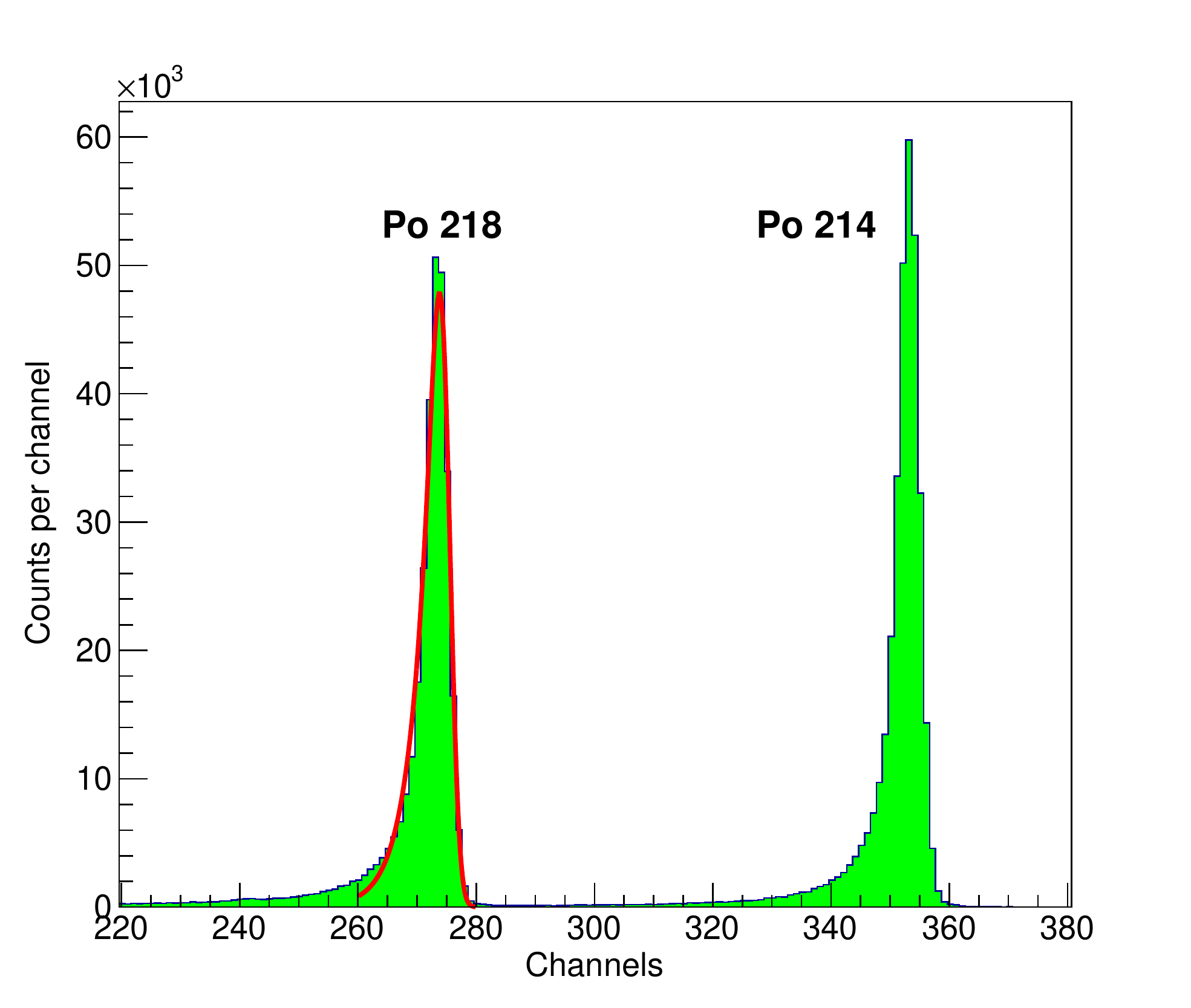}
\caption{A typical pulse height radon spectrum with $^{218}$Po, fitted to function (1), and $^{214}$Po alpha-peaks with kinetic energies of 6.00 MeV and 7.69 MeV, respectively.} 
\end{figure}%

\begin{figure}[h!]%
        \centering
        \includegraphics[height=3in]{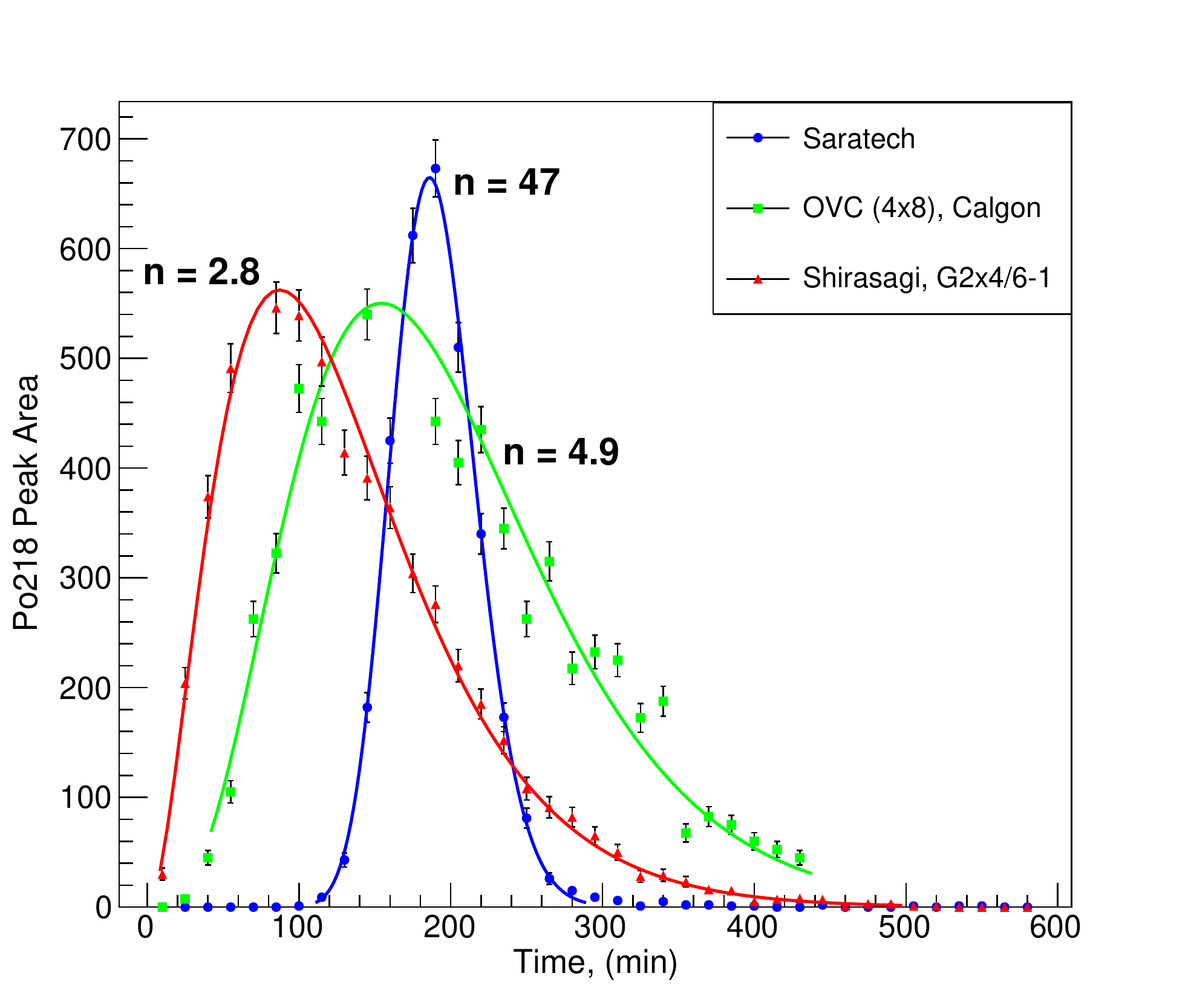}
        \caption{Examples of elution curves, fitted to function (2) (solid lines), of $^{218}$Po measured in Ar carrier gas at 295 K where n is the number of theoretical stages.} 
\end{figure}%

The chromatographic plate model method \cite{Levins1979, Keulemans1959} was employed in the data analysis for this paper. In this approach, a charcoal trap can be divided into a number of stages of equal volume in which equilibrium can always exist between the gas and the charcoal. For a short pulse of $^{222}$Rn introduced into the charcoal bed layer the elution curve is

\begin{equation}
y\big(\frac{t}{\tau}\big)=\frac{\alpha n^n}{(n-1)!}\big(\frac{t}{\tau}\big)^{n-1}e^{-n \frac{t}{\tau}},
\end{equation}

where $\alpha$ is the amplitude of the radon spike input, $\tau$ is the average breakthrough time of $^{222}$Rn in a carrier gas, and $\textit{n}$ is the number of theoretical stages.

The $\tau$-value is given by the linear relation  \cite{Pocar2003} 

\begin{equation}
\tau=\frac{k_a m}{f},
\end{equation}

where $\textit{$k_a$}$  is the dynamic adsorption coefficient in l/g, $\textit{m}$ is the mass of the adsorbent in g, and $\textit{f}$ is the mass flow rate in slpm.

The $\tau$-values for $^{222}$Rn in N$_2$, Ar and Xe gases and the $\textit{n}$-values were obtained from fits of the elution curves and are presented with total uncertanties ($\delta$$\tau$$^2$=$\sigma$$^2$$_{stat}$+$\sigma$$^2$$_{sys}$) in Table 3. The $\textit{n}$-values are presented with statistical uncertanties derived from the elution curve fits.


\begin{sidewaystable}
\resizebox{1.0\textwidth}{!}{
\centering
\begin{tabular}{|c|c|c|c|c|c|c|c|c|c|c|c|}
\hline
\multirow{2}{*}{Carrier Gas} & \multirow{2}{*}{Charcoal Type} & \multicolumn{5}{c|}{$\tau$(min)}                                                                                                                                           & \multicolumn{5}{c|}{n}                                                                            \\ \cline{3-12}
                                                                            &                                & 295 K           & 273 K           & 263 K           & 253 K           & 190 K & 295 K           & 273 K           & 263 K           & 253 K        & 190 K \\ \hline
\multirow{3}{*}{\begin{tabular}[c]{@{}c@{}}N$_2$\\ (2 slpm)\end{tabular}} & Calgon OVC 4x8 (50 g, 1 AtmA)                    & 174.0$\pm$7.5            & 419$\pm$19                 & 712$\pm$31                   & 1123$\pm$50        & $-$           & 4.7$\pm$0.1         & 5.4$\pm$0.2         & 5.5$\pm$0.2         & 5.6$\pm$0.2         & $-$  \\ \cline{2-12} 
                                                                            & Saratech (70 g, 1 AtmA)                       & 189$\pm$8                 & 444$\pm$20                 & 709$\pm$31                  & 1152$\pm$50                  &$-$          & 46$\pm$1                 & 52$\pm$1                 & 62$\pm$2                 & 61$\pm$1                 &$-$   \\ \cline{2-12} 
                                                                            
& Shirasagi (45 g, 1 AtmA) & 123$\pm$6 & 309$\pm$15 & 490$\pm$23 & 825$\pm$39 & $-$ & 3.10$\pm$0.02 & 3.3$\pm$0.1 & 3.5$\pm$0.1 & 3.7$\pm$0.1 & $-$
\\ \cline{2-12} 
\hline
\multirow{3}{*}{\begin{tabular}[c]{@{}c@{}}Ar\\ (2 slpm)\end{tabular}} & Calgon OVC 4x8 (50 g, 1 AtmA)                   & 210$\pm$9                 & 505$\pm$23                 & 825$\pm$36                   & 1037$\pm$47                 &$-$           & 5.0$\pm$0.2         & 5.0$\pm$0.2         & 5.4$\pm$0.3         & 6.2$\pm$0.4         & $-$  \\ \cline{2-12} 
                                                                            & Saratech (70 g, 1 AtmA)                       & 189$\pm$9                 & 520$\pm$24                 & 844$\pm$38                  & 1371$\pm$62                  &$-$          & 47$\pm$1                 & 52$\pm$1                 & 58$\pm$2                & 71$\pm$2                 &$-$   \\ \cline{2-12} 
                                                                            
& Shirasagi (45 g, 1 AtmA) & 117$\pm$6 & 366$\pm$17 & 613$\pm$31 & 1015$\pm$57 & $-$ & 3.3$\pm$0.1 & 3.0$\pm$0.1 & 3.1$\pm$0.1 & 3.0$\pm$0.1 & $-$
\\ \hline

\multirow{3}{*}{\begin{tabular}[c]{@{}c@{}}Xe\\ (0.5 slpm)\end{tabular}} & CarboAct (241 g, 1 AtmA)                    & 246$\pm$14                 & 329$\pm$18                 & 407$\pm$22                   & 563$\pm$31                 &992$\pm$54           & 8.3$\pm$0.5         & 8.6$\pm$0.2         & 10.3$\pm$1.0         & 5.9$\pm$0.1         & 5.60$\pm$0.04  \\ \cline{2-12} 
                                                                            & Saratech (650 g, 1.6 AtmA)                       & 612$\pm$32                 & 800$\pm$42                 & 1270$\pm$67                  & 1651$\pm$88                  &3800$\pm$200          & 138$\pm$2                 & 132$\pm$1                 & 142$\pm$1                 & 142$\pm$10                 &142$\pm$9   \\ \cline{2-12} 
                                                                            
& Etched Saratech (650 g, 1.6 AtmA) & 623$\pm$33 & 882$\pm$47 & 1216$\pm$64 & 1788$\pm$95 & 3840$\pm$200 & 117$\pm$3 & 132$\pm$5 & 116$\pm$5 & 140$\pm$11 & 139$\pm$3
\\ \cline{2-12} 
                                                                            & Shirasagi (45 g, 1 AtmA)                      & 46$\pm$3                  & 63$\pm$4                  & 85$\pm$5                  & 109$\pm$7                 & 193$\pm$12          & 6.0$\pm$0.1                   & 5.70$\pm$0.04                   & 5.50$\pm$0.05                   & 5.7$\pm$0.1                   &  4.2$\pm$0.1  \\ \hline
\begin{tabular}[c]{@{}c@{}}Xe (2 slpm)\end{tabular} & Shirasagi (45 g, 1 AtmA) & 13.8$\pm$0.1 & 14$\pm$1 & 20$\pm$1 & 29$\pm$2 & 50$\pm$3 & 2.90$\pm$0.04 & \multicolumn{1}{l|}{2.90$\pm$0.04} & \multicolumn{1}{l|}{2.00$\pm$0.02} & \multicolumn{1}{l|}{3.0$\pm$0.1} & \multicolumn{1}{l|}{2.30$\pm$0.04}  \\ \hline
\end{tabular}%
}
\caption{The average $^{222}$Rn breakthrough times and number of theoretical stages in various types of charcoals measured in N$_2$, Ar and Xe carrier gases. The $\tau$-values include total measurement uncertainties.}
\label{my-label}
\end{sidewaystable}

The average breakthrough times of $^{222}$Rn measured in Ar carrier gas in all charcoals are greater than in N$_2$ gas measured in the same charcoals. The longest average breakthrough times of $^{222}$Rn both in N$_{2}$ and Ar carrier gases were measured in Calgon OVC 4x8 and Saratech. The total uncertainties were about 5\%. The systematic uncertainties were due to small fluctuations in the flow rate and gas pressure in the system. The $\tau$-values in Ar and N$_2$ in Calgon OVC 4x8 and Saratech charcoals increased with increasing gas pressure. 

The $\tau$-values for $^{222}$Rn in Xe measured in Saratech are greater than in Carboact. The $\tau$-values for regular Saratech and HNO$_3$ etched Saratech are consistent within statistical and systematic uncertainties. The $\tau$-values in Xe gas in Carboact and Shirasagi decrease as the gas flow rate increases by a factor of four. The total uncertainties were about 5\% as well. Unlike in Ar and N$_2$ carrier gases, no difference in the $\tau$-values was observed in Xe carrier gas when the gas pressure was increased. The measurements of $^{222}$Rn breakthrough times in carrier gases in various charcoals allowed calculating $^{222}$Rn dynamic adsorption coefficients $\textit{$k_a$}$ using relationship (3), which determine optimal parameters for charcoal traps. The $\textit{$k_a$}$-values of $^{222}$Rn in N$_2$, Ar and Xe gases were calculated for various charcoals and are presented in Table 4.


\begin{sidewaystable}
\centering
\begin{tabular}{ |c|c|c|c|c|c|c|c|c|}
	\hline                                  
	\multicolumn{8}{|c|} {$\textit{$k_a$, l/g}$} \\
	\hline
	Carrier gas & Charcoal brand & 295 K & 273 K & 268 K & 263 K & 258 K & 253 K \\ \hline	
	N$_2$ (2 slpm, 1 atmA) & Calgon OVC 4x8 (50 g) & 7.0$\pm$0.3 & 17$\pm$1 & 23$\pm$1 & 29$\pm$1 & 34$\pm$2 & 45$\pm$2 \\ \hline
	N$_2$ (2 slpm, 1 atmA) & Saratech (70 g) & 5.4$\pm$0.3 & 13$\pm$1 & $-$ & 20.2$\pm$1.0 & $-$ & 33$\pm$2 \\ \hline
	N$_2$ (2 slpm, 1 atmA) & Shirasagi (45 g) & 5.5$\pm$0.3 & 14$\pm$1 & $-$ & 22$\pm$1 & $-$ & 37$\pm$2 \\ \hline
	Ar (2 slpm, 1 atmA) & Calgon OVC 4x8 (50 g) & 8.4$\pm$0.4 & 20$\pm$1 & 25$\pm$1 & 33$\pm$2 & 42$\pm$2 & $-$ \\ \hline
	Ar (2 slpm, 1 atmA) & Saratech (70 g) & 5.4$\pm$0.3 & 15$\pm$1 & $-$ & 24$\pm$1 & $-$ & 39$\pm$2 \\ \hline
	Ar (2 slpm, 1 atmA) & Shirasagi (45 g) & 5.2$\pm$0.3 & 16$\pm$1 & $-$ & 27.2$\pm$1.4 & $-$ & 45$\pm$3 \\  \hline \hline
	Carrier gas & Charcoal brand & 295 K & 273 K & 253 K & 233 K & 213 K & 190 K \\ \hline 
	Xe (0.5 slpm, 1 atmA) & Carboact (241 g) & 0.51$\pm$0.02 & 0.68$\pm$0.03 & 0.84$\pm$0.04 & 1.17$\pm$0.05 & $-$ & 2.05$\pm$0.10 \\ \hline
	Xe (0.5 slpm, 1.6 atmA) & Saratech (650 g) & 0.50$\pm$0.03 & 0.61$\pm$0.03 & 1$\pm$0.1 & 1.3$\pm$0.1 & 2.0$\pm$0.1 & 2.9$\pm$0.2 \\ \hline
	Xe (0.5 slpm, 1.6 atmA) & Etched Saratech (650 g) & 0.50$\pm$0.03 & 0.70$\pm$0.04 & 1.0$\pm$0.1 & 1.4$\pm$0.1 & 2.0$\pm$0.1 & 3.0$\pm$0.2 \\ \hline
	Xe (0.5 slpm, 1 atmA) & Shirasagi (45 g) & 0.51$\pm$0.03 & 0.70$\pm$0.04 & 1.0$\pm$0.1 & 1.2$\pm$0.1 & $-$ & 2.1$\pm$0.1 \\ \hline
	Xe (2 slpm, 1 atmA) & Carboact (241 g) & 0.44$\pm$0.01 & 0.73$\pm$0.02 & 0.94$\pm$0.02 & 1.22$\pm$0.03 &  1.62$\pm$0.04 &  2.1$\pm$0.1 \\ \hline
	Xe (2 slpm, 1 atmA) & Shirasagi (45 g) & 0.61$\pm$0.03 & 0.90$\pm$0.04 & 1$\pm$0.1 & 1.3$\pm$0.1 & $-$ & 2.2$\pm$0.1 \\ \hline
\end{tabular}
\caption{Dynamic adsorption coefficients of $^{222}$Rn on various charcoals in N$_2$, Ar and Xe carrier gases. The values include total measurement uncertainties.}
\label{my-label}
\end{sidewaystable}

As shown in Table 4, the $^{222}$Rn adsorption coefficients range from 5 to 45 l/g in N$_2$ and Ar carrier gases depending on the charcoal and its temperature. The $\textit{$k_a$}$-values for $^{222}$Rn obtained in Xe carrier gas are about an order of magnitude lower and fall in the range of 0.5-3 l/g. The plots of $^{222}$Rn adsorption coefficients for various charcoals in N$_2$, Ar and Xe carrier gases are shown in Figures 6 and 7 (a,b).

\begin{figure}[h!]%
        \centering
        \includegraphics[height=2.7in]{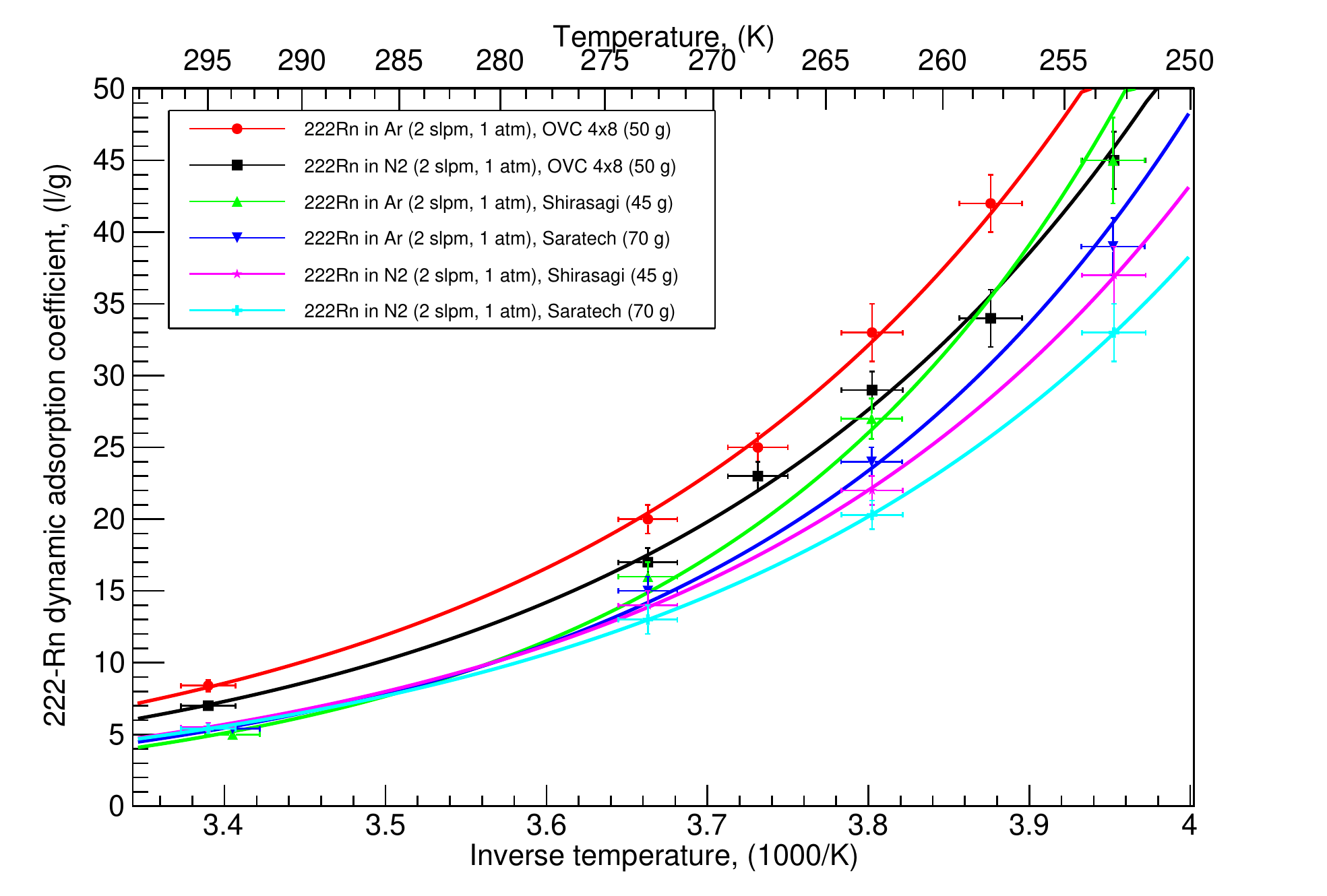}
        \caption{Dynamic adsorption coefficients with total measurement uncertainties, fitted to the Arrhenius equation (solid lines), in various charcoals measured in N$_{2}$ and Ar carrier gases vs inverse temperature.} 
\end{figure}%

\begin{figure}[t]
 \centering
 \subfloat[]{\includegraphics[width=0.55\textwidth]{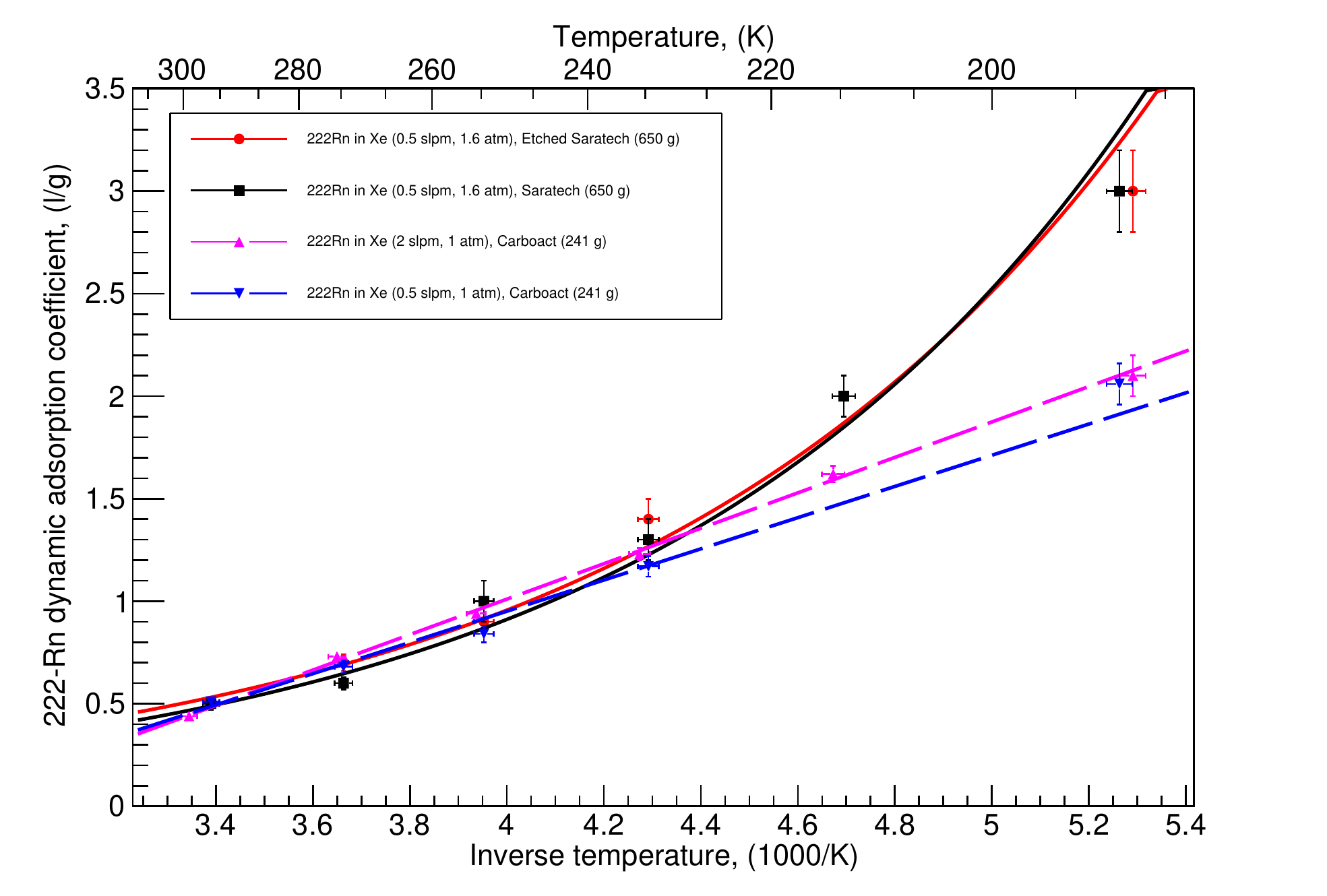}\label{fig:f1}}
 \vspace{0.00001cm}
 \hfill
 \subfloat[]{\includegraphics[width=0.55\textwidth]{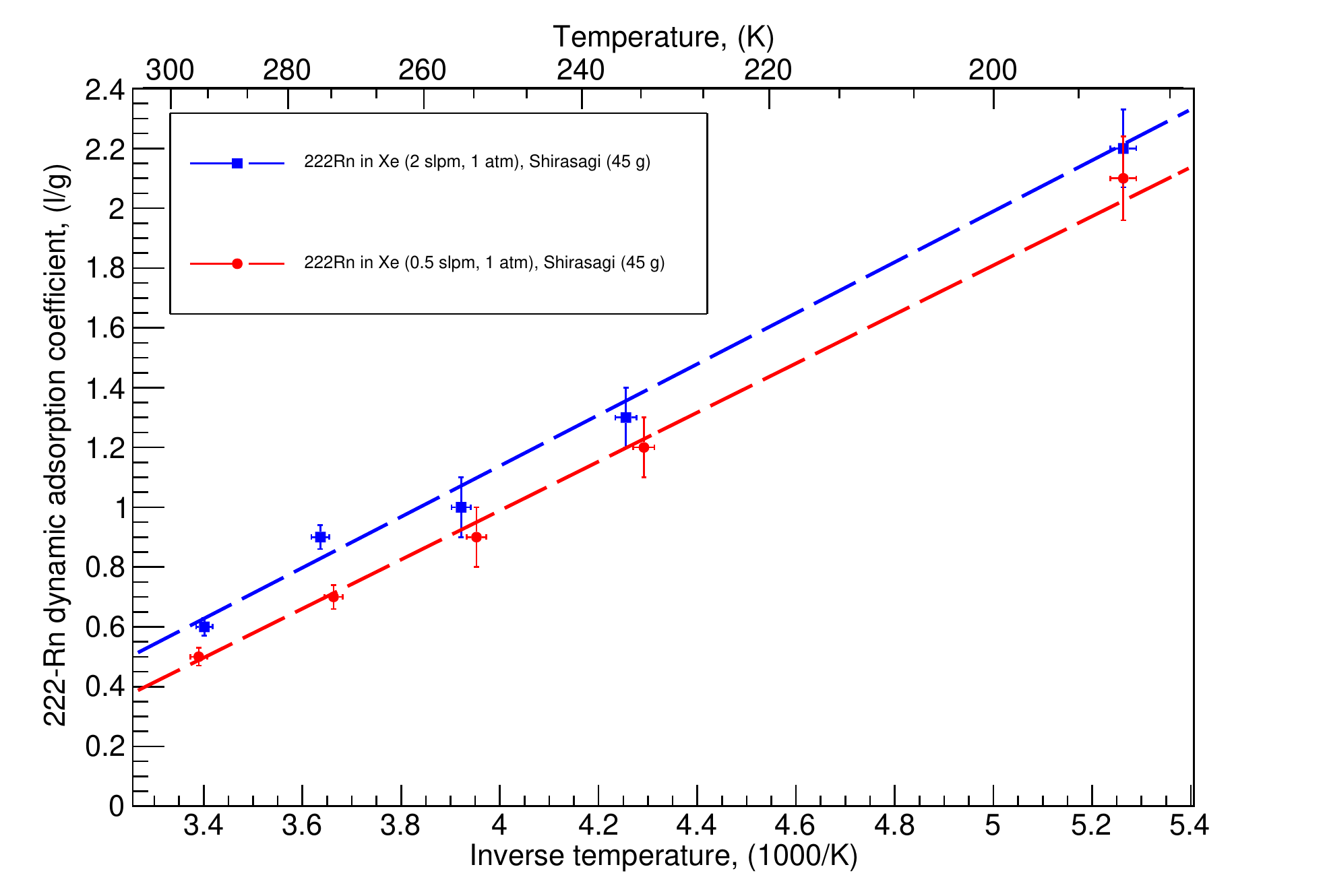}\label{fig:f2}}
 \caption{Dynamic adsorption coefficients with total measurement uncertainties, fitted to the Arrhenius equation (solid lines) measured in Xe carrier gas vs inverse temperature in a) regular and HNO$_3$ etched Saratech, Carboact (dashed lines, linear fit) and in b) Shirasagi (dashed lines, linear fit).}
\end{figure}



The $\textit{$k_a$}$-values were fitted to the Arrhenius equation \cite{Gross}  as a function of inverse temperature in Ar and N$_2$ carrier gases as shown in Figure 6.
The Arrhenius relation, shown below, gives the dependence of the $\textit{$k_a$}$-value for $^{222}$Rn atoms adsorbed on charcoals where $\textit{k$_0$}$ is the pre-exponential factor, frequency in (1/s), that yields the numbers of attempts by a particle to overcome a potential barrier, $\textit{$Q$}$ is the adsorption heat (J/mol), $\textit{R}$ is the universal gas constant in (J/mol*K), and $\textit{T}$ is the absolute temperature in (K).

\begin{equation}
k_a=k_0e^{\frac{Q}{RT}}.
\end{equation}

The $\textit{$k_a$}$-values are greater in Ar gas than in N$_2$ gas, and they are greater for Calgon OVC 4x8 charcoal than for Shirasagi and Saratech. Figure 7 (a,b) shows that the $^{222}$Rn dynamic adsorption coefficients in Xe carrier gas for regular Saratech and HNO$_3$ etched Saratech are consistent within statistical and systematic uncertanties. While the $\textit{$k_a$}$-values appear to obey the Arrhenius relationship for regular and HNO$_3$ etched Saratech, they violate it for Carboact and Shirasagi and show a linear relation as a function of temperature. 

\subsection{Discussion}

The measurements of the $\tau$ and $\textit{$k_a$}$-values are crucial for demonstrating the behavior of $^{222}$Rn atoms in various charcoals. They revealed that both the $\tau$ and $\textit{$k_a$}$-values are significantly greater in N$_{2}$ and Ar carrier gases than in Xe gas. This effect may be attributed to the low polarizabilities of N$_{2}$ and Ar gases which leads to their low attraction to charcoals. The adsorption of N$_{2}$ and Ar gases was measured in the charcoals in the 0.1 l trap during the $^{222}$Rn adsorption characteristics measurements in the range of 253-295 K. The adsorbed mass of N$_{2}$ and Ar was below the detection limit of the scale. In contrast, Xe atoms have high polarizability and tend to occupy the charcoal adsorption sites almost instantly resulting in short $^{222}$Rn breakthrough times \cite{Boer1957}. The adsorbed mass of Xe, scaled to 1 kg of Saratech and Carboact, as a function of temperature is shown in Figure 8. The xenon adsorption measurements in Saratech were crosschecked using tension load cells and the results agreed within statistical and systematic uncertainties. The adsorbed mass of Xe increases linearly in Carboact and Saratech with decreasing temperature. The adsorbed mass of Xe in Carboact did not rise proportionally with increasing gas pressure. It is evident from Figure 8 that Saratech adsorbs on average 30\% less Xe than Carboact at atmospheric pressure. In order to measure carrier gases adsorption and their interference in $^{222}$Rn adsorption on charcoals with high precision, mass-spectroscopy methods should be applied.  

\begin{figure}[t]%
        \centering
        \includegraphics[height=2.7in]{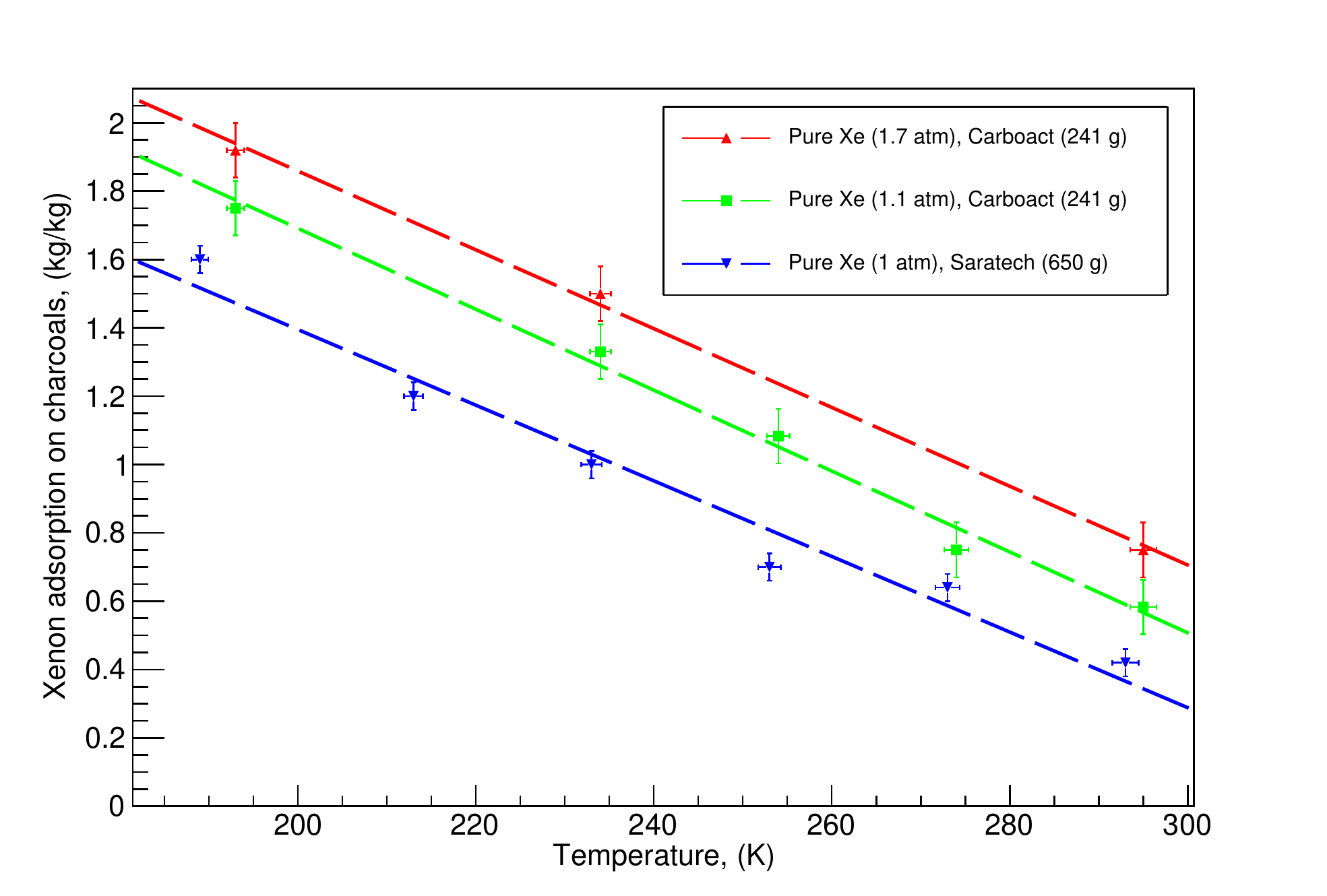}
        \caption{Adsorption of Xe gas in 1 kg of charcoal with total measurement uncertainties, fitted to linear fit (dashed lines), vs temperature.} 
\end{figure}%

The linear dependence of the $\textit{$k_a$}$-values shown in Figure 7 (a,b) for Carboact and Shirasagi may be attributed to saturation effects of the charcoal's surface area by Xe atoms as a function of temperature. The $^{222}$Rn dynamic adsorption coefficients on charcoal in Ar gas are greater than in N$_2$ gas. 
\par
The shapes of the $^{222}$Rn elution curves determine the operation of the adsorbing bed of charcoals in the trap which is another important characteristic. It is worthwhile pointing out that for $\textit{$n$}$$\rightarrow$$\infty$, the elution curve becomes symmetric and the $\tau$-values tend to approach a gaussian distribution with a standard deviation of 

\begin{equation}
\sigma{_{\tau}}=\frac{\tau}{\sqrt n}.
\end{equation}

Figure 5 shows that the shape of the elution curve measured in Saratech is more symmetric than the curves obtained for Shirasagi and Calgon OVC 4x8. The shapes of the elution curves are independent of the used carrier gases. It is possible that the shape of the elution curves may be affected by the shape of the charcoal granules. 

\begin{figure}[t]
 \centering
 \subfloat[]{\includegraphics[width=0.2\textwidth]{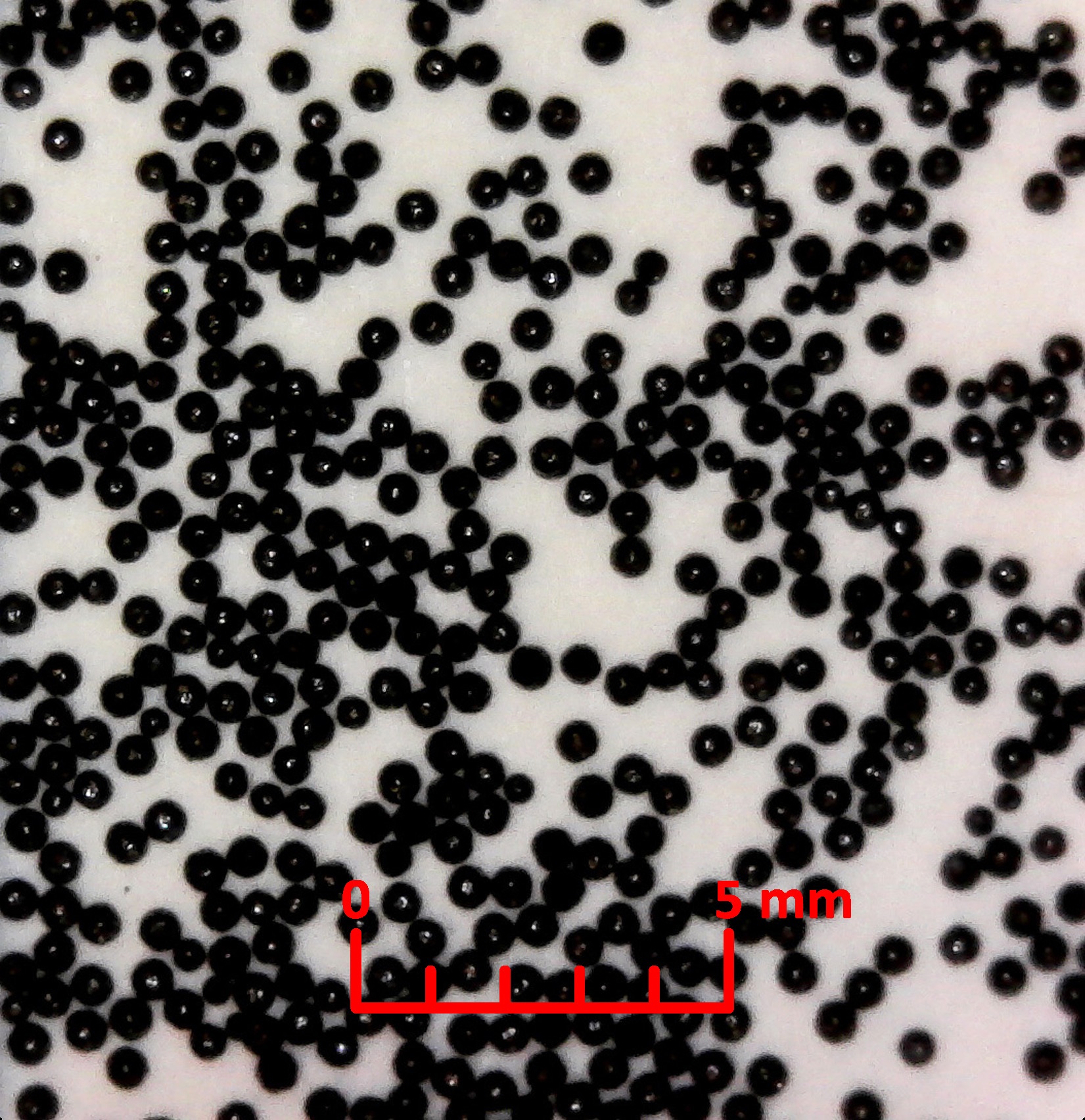}\label{fig:f1}}
 \hfill
 \subfloat[]{\includegraphics[width=0.2\textwidth]{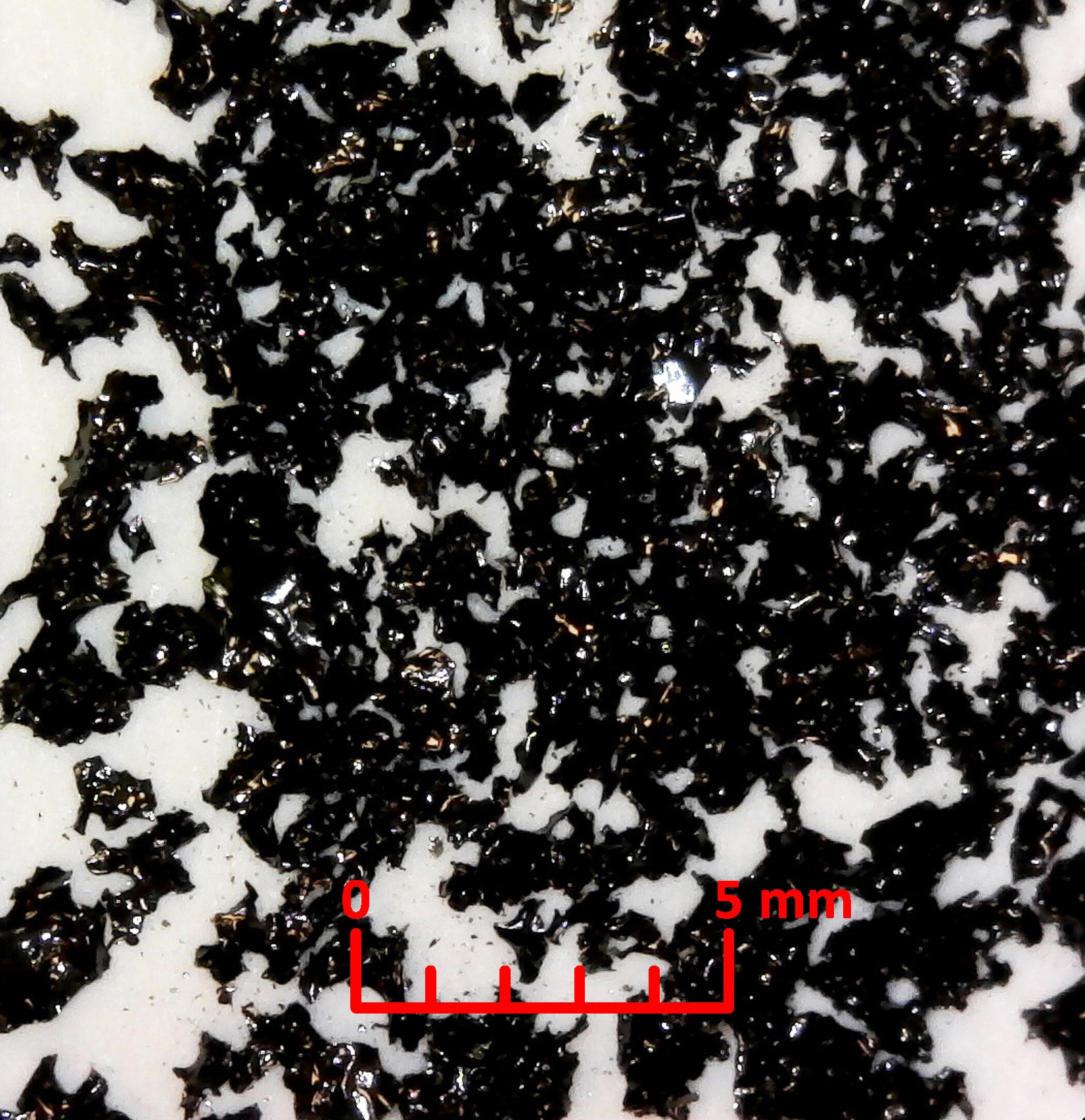}\label{fig:f2}}
 \hfill
 \subfloat[]{\includegraphics[width=0.2\textwidth]{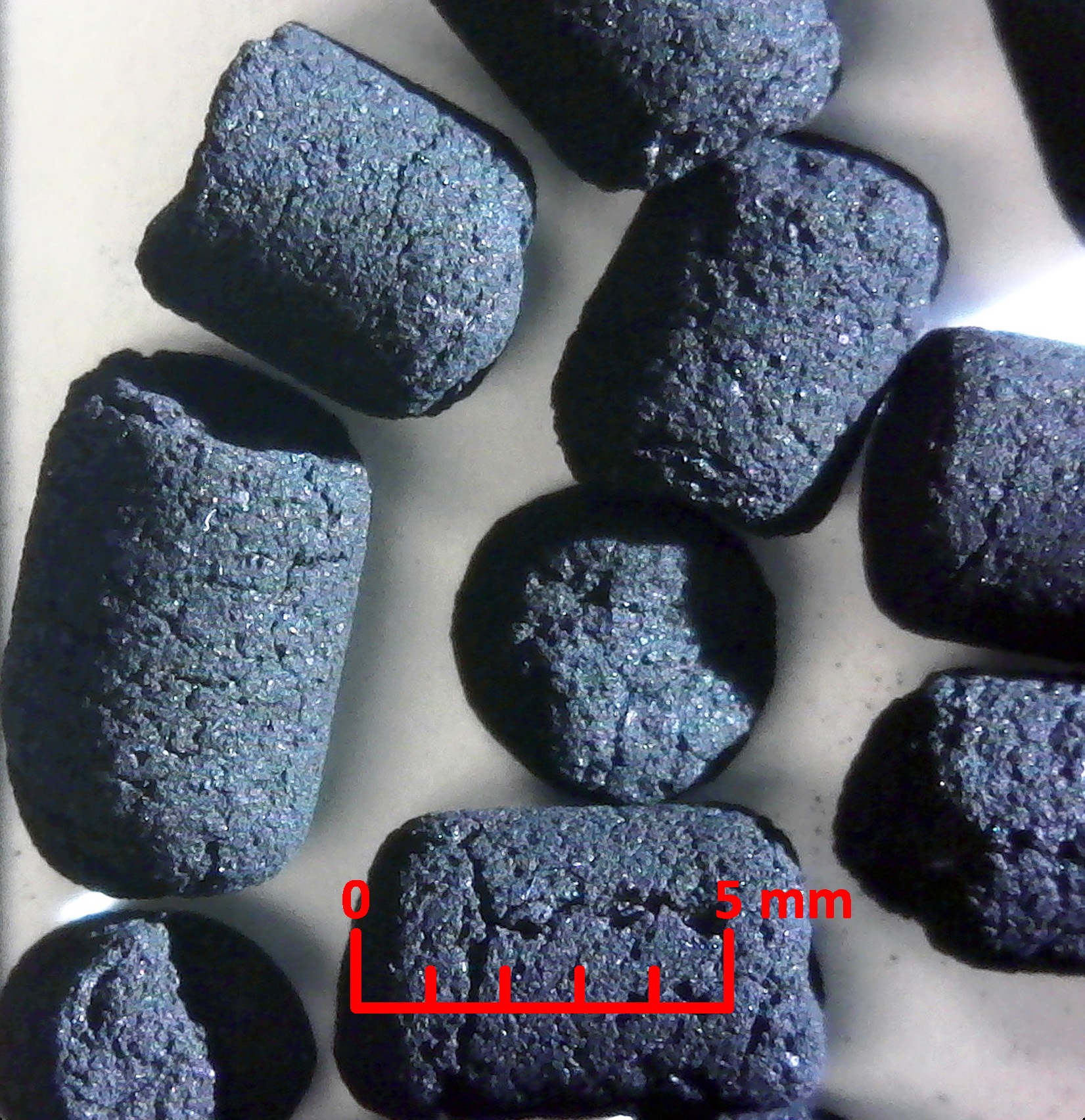}\label{fig:f3}}
 \caption{Microscopic images of Saratech spherical adsorbent with an average diameter of 0.5 mm (a), Carboact adsorbent with nonuniform, fragmented shape and size (b) and Shirasagi adsorbent with nonuniform, cylindrical shape (c).}
\end{figure}

Microscopic images of Saratech, Carboact and Shirasagi shown in Figure 9 reveal the large difference in shape and size for the charcoal granules. The average shape and size of Saratech granules are uniform compared to Carboact, Calgon OVC 4x8 and Shirasagi granules. It will lead to less open spaces between the Saratech granules resulting in a significant increase of the total surface area of the adsorbent to the volume of the trap \cite{Ruthven}. As a consequence, the number of theoretical stages, $\textit{$n$}$, is higher for Saratech than for Carboact, Shirasagi, and Calgon OVC 4x8 as shown in Table 3. This defines the uncertainty of the average $^{222}$Rn breakthrough time pulse through the charcoal trap, i.e. the lower the $\textit{$n$}$-value is the sooner $^{222}$Rn atoms begin departing the charcoal trap despite the fact that the average breakthrough time remains the same. Hence, this study of $^{222}$Rn adsorbing characteristics have shown that Saratech, among all the investigated charcoals, appears to be the most efficient $^{222}$Rn adsorbent. Moreover, the chemical treatment of Saratech leads to a significant reduction of radioactivity while it retains its $^{222}$Rn adsorbing properties. This makes it particularly desirable and cost effective for $^{222}$Rn reduction applications in low background experiments. 
\par



\subsection{Trap performance}

Equation (6) predicts the efficacy of a charcoal trap to reduce $^{222}$Rn concentration in TPC detectors based on the total mass and specific activity of a charcoal material according to

\begin{equation}
N_{out}=N_{in}e^{-\frac{k_a \cdot m}{f\cdot \tau_R}}+s_0\, f\, \frac{\tau_R}{k_a}\left(1-e^{-\frac{k_a \cdot m}{f\cdot \tau_R}}\right),
\end{equation}

where $\tau_R$ is the average $^{222}$Rn lifetime (7921 min), $s_0$ is the $^{222}$Rn specific activity in mBq/kg, $\textit{N$_{in}$}$ is the $^{222}$Rn concentration entering the charcoal trap in mBq, and $\textit{N$_{out}$}$ is the total $^{222}$Rn concentration at the output of the charcoal trap in mBq. It should be noted that the lowest achievable $^{222}$Rn concentration at the output of the trap is given by 
$min(N_{out})=s_0f\tau_R/k_a$, and thus depends on the specific activity but not on the total mass of the charcoal.

Figure 10 provides an illustration of Equation (6) for HNO$_3$ etched Saratech and Carboact for input $^{222}$Rn concentrations of (a) 8.3 mBq and (b) 20 mBq entrained in Xe carrier gas both at a flow rate of 0.5 slpm, respectively. The $^{222}$Rn concentration range corresponds to the estimated concentrations that continuously emanate from the warm LZ detector components embedded in xenon gas and are used for the inline radon reduction system being constructed for the LZ DM search experiment. The radon reduction results suggest that it will require about 5(7) kg of etched Saratech to reduce the estimated $^{222}$Rn concentrations of 8.3(20) mBq  \cite{Akerib2018, LZTDR}, continually emanated from the LZ detector components, below 1 mBq in the return stream of the radon reduction system. The dynamic adsorption coefficients, used in these calculations for etched Saratech and Carboact, were measured at 190 K. 
For fixed and relatively limited volumes of adsorbent, the much higher density for Saratech produces more efficient retention of radon at a cost that is about 50 times lower than for Carboact. Thus, etched Saratech provides a very attractive option for an effective inline radon reduction system based on excellent performance and low cost.

\begin{figure}[ht!]
 \centering
 \subfloat[]{\includegraphics[width=0.55\textwidth]{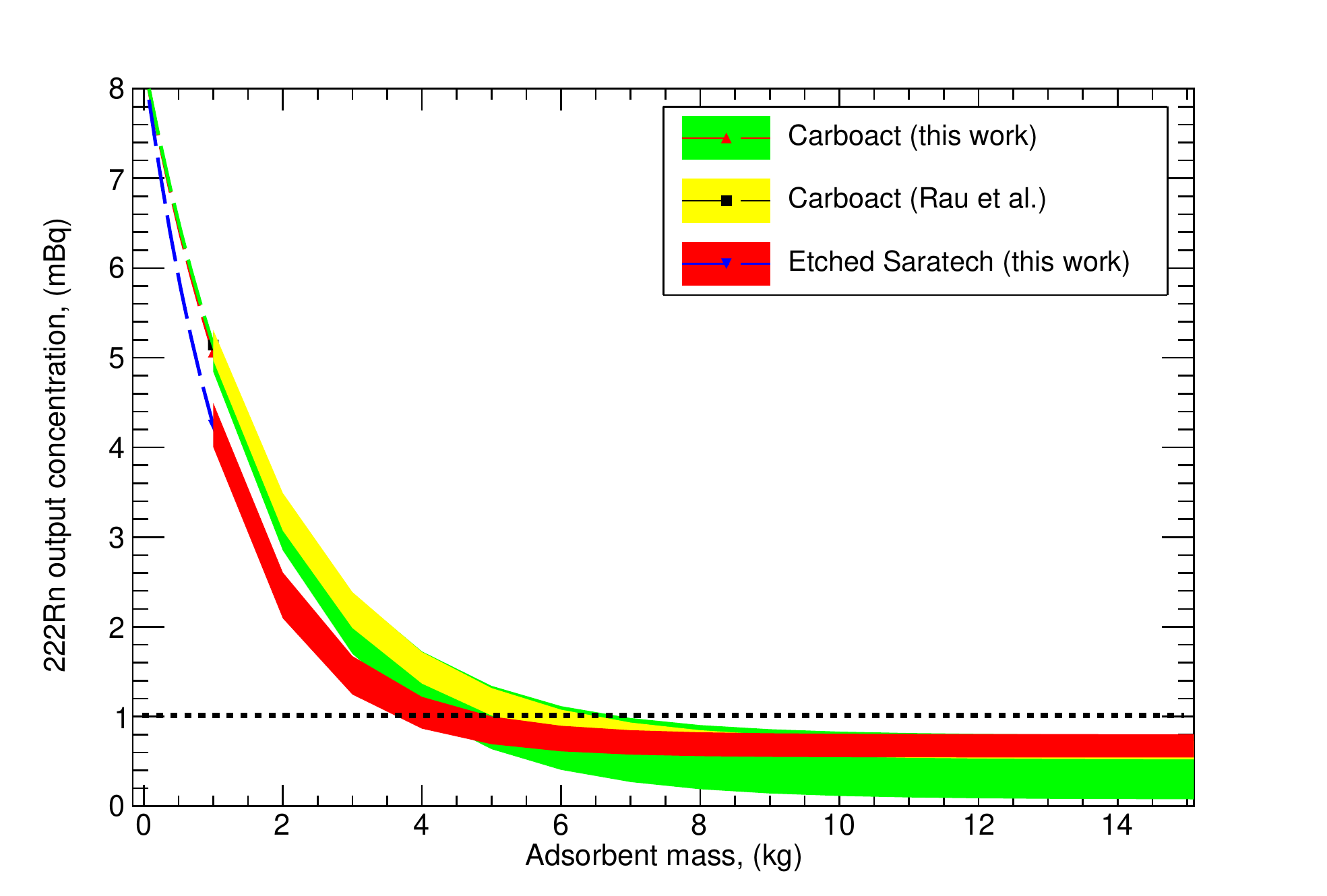}\label{fig:f1}}
 \hfill
 \subfloat[]{\includegraphics[width=0.55\textwidth]{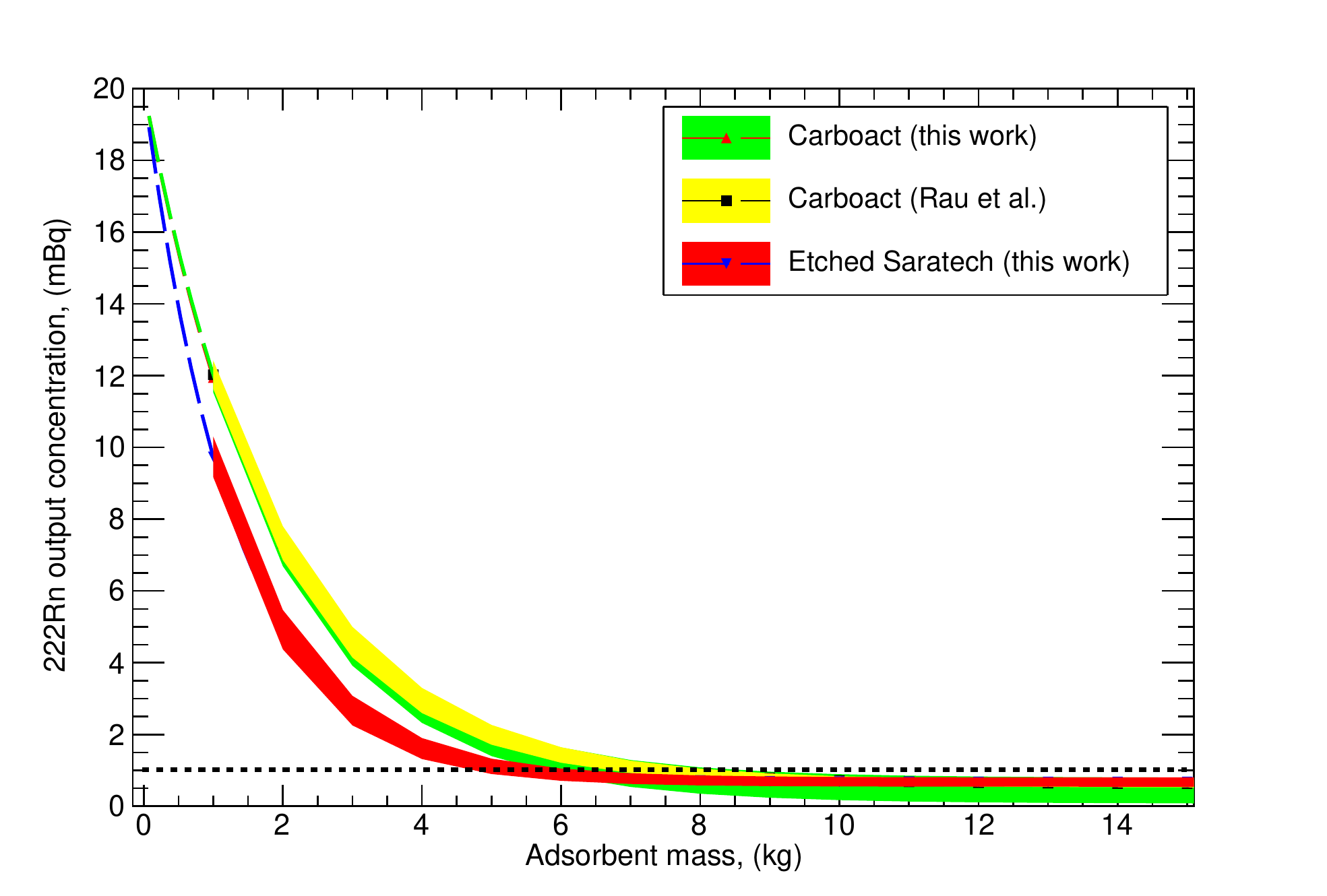}\label{fig:f2}}
 \caption{$^{222}$Rn trap output concentration versus charcoal mass for input trap concentrations of 8.3 mBq (a), and 20 mBq (b) calculated both for the Xe flow rate of 0.5 slpm for the LZ DM detector. The colored bands represent the quadrature sums of the statistical and systematic uncertainties, and the black dashed lines represent the minimal $^{222}$Rn concentrations required for the LZ DM experiment. The dynamic adsorption coefficients for Etched Saratech and Carboact used in the calculation were measured at 190 K.}
\end{figure}

\section{Conclusion}

$^{222}$Rn adsorbing characteristics in charcoals were measured in N$_{2}$, Ar, and Xe carrier gases in the temperature range of 190-295 K at different gas flow rates. The measurements have shown that breakthrough times of $^{222}$Rn entrained in N$_{2}$ and Ar carrier gases are significantly longer than in Xe carrier gas. This effect may be attributed to the low polarizabilities of N$_{2}$ and Ar gases requiring significantly smaller amounts of charcoals to effectively trap $^{222}$Rn. In contrast, $^{222}$Rn atoms in Xe have much shorter average breakthrough times and much smaller dynamic adsorption coefficients due to saturation effects of Xe atoms in charcoals resulting in far fewer available adsorption sites for $^{222}$Rn atoms. The adsorption measurements of Xe gas in various charcoals have revealed that Saratech adsorbs Xe about 30\% less than Carboact at atmospheric pressure independent of temperature. Both the $\tau$ and $k_a$-values for $^{222}$Rn in Ar and N$_2$ carrier gases follow the Arrhenius law which describes adsorption and desorption kinetic processes on surfaces. However, for $^{222}$Rn in Xe carrier gas this is only true for Saratech, but not for Carboact and Shirasagi, where $\tau$ and $k_a$ do not follow the Arrhenius law, but instead display a linear dependence as a function of inverse temperature. This may be attributed to higher saturation processes. Among all investigated charcoals, Saratech appears to be the most efficient $^{222}$Rn reduction material. The chemical treatment of Saratech with ultra-pure HNO$_3$ acid reduced its intrinsic radioactivity ($^{238}$U) and, as a consequence, the $^{222}$Rn specific activity by a factor of three and made it competitive with Carboact. Moreover, the etching did not affect the $^{222}$Rn adsorption characteristics making Saratech a strong candidate for the future DM and NDBD low background experiments considering its very low cost and exemplary properties.

\section{Acknowledgments}
We acknowledge support of the US Department of Energy (DE-SC0015708), Lawrence Berkeley National Laboratory (LBNL), Stanford Linear Accelerator Center (SLAC) and the University of Michigan. We would like to thank Professor Jacques Farine at Laurentian University for providing the rubber radon emanation standard and Professor Tom Shutt at SLAC for providing the radon counting hardware. We would like to acknowledge Eric Miller at South Dakota School of Mines and Technology for performing the temperature dependence radon emanation rate calculations for the LZ components. We would also like to thank the members of the LZ collaboration for many insightful discussions. Special thanks go to Professor Kimberlee Kearfott at the Nuclear Engineering and Radiological Sciences department of the University of Michigan for lending us the Pylon radon source.


\end{document}